\documentclass[12pt,twoside,pra,aps,amssymb,amsfonts,amsmath,tightenlines,
nofootinbib]{revtex4}

\usepackage{amssymb}
\usepackage{amsmath}
\usepackage{amsfonts}
\usepackage{amsbsy}
\usepackage{amstext}
\usepackage{dsfont}
\usepackage{graphicx}
\usepackage[caption = false]{subfig}
\usepackage{floatrow}
\newcommand{\half}{\mbox{$\textstyle \frac{1}{2}$}}

\newcommand{\re}{\mbox{$\rm e$}}
\newcommand{\ri}{\mbox{$\rm i$}}
\newcommand{\rd}{\mbox{$\rm d$}}
\begin{document} 

\title{Quantum clock and Newtonian time}

\author{Dorje C. Brody$^{1,2}$ and Lane P. Hughston$^{3,4}$}

\affiliation{
$^{1}$School of Mathematics and Physics, University of Surrey, Guildford 
GU2 7XH, UK \\ 
$^{2}$Department of Mathematics, Imperial College London, London SW7 2BZ, UK  \\ 
$^{3}$School of Computing, Goldsmiths University of London, New Cross,
London SE14 6NW, UK\\
$^{4}$Artificial Intelligence and Mathematics Research Lab, James Carter Road, Mildenhall, Bury St Edmunds IP28 7DE, UK}

\begin{abstract}
\noindent 
An extension of standard quantum mechanics is proposed in which the Newtonian time appearing as a parameter in the unitary evolution operator is replaced with the time shown by a `quantum clock'. Such a clock is defined by the following properties:  (a) the time that the clock shows is nondecreasing, (b) the clock ticks at random Newtonian times with random tick sizes, and (c) on average the clock shows the Newtonian time. We show that the leading term in the evolution equation for the density matrix associated with any quantum clock  gives the von Neumann equation. The leading correction to the von Neumann equation is given by the Lindblad equation generated by the Hamiltonian, but there are higher-order terms that generalize the von Neumann equation and the Lindblad terms. Modifications to the von Neumann equation are worked out in detail in a parametric family of models for which the tick sizes are gamma distributed. Lower bounds on the parameters of these quantum clock models are derived using the precision limit of an atomic clock. An anomalous term in the Ehrenfest theorem for a free particle is derived, which in principle can be used as a basis for testing such models.
\end{abstract}

\maketitle
 
\noindent In quantum mechanics,  the time evolution of the state of a system is 
formulated on the basis of a notion of time firmly grounded in Newtonian 
physics -- in which time is a fixed structure external to the system \cite{Isham}. 
Yet, empirically, the timescales of physical processes, especially in the 
microscopic realm, appear to be dependent on the nature of the environment to 
which the system is exposed. For example, chemical reactions can vary from 
femtoseconds to years, depending on a range of factors including the 
concentration and density of the reactants, the pressure, the temperature, and 
the presence of catalysts. At high concentration, a process can move forward 
rapidly to completion, whereas in a sparse environment, a process progresses 
more slowly. If a system engages in many interactions with its environment, then 
time in effect passes more quickly for the system; whereas for a nearly isolated 
system with little by way of  interaction, time passes slowly. 

This observation motivates us to explore whether it is reasonable to assume that 
the Newtonian clock process $\{t\}_{t \geq 0}$ subordinating the unitary evolution 
operator $\{{\hat U}(t)\}_{t\geq0}$ governing the dynamics of a quantum system is 
necessarily deterministic. Our purpose here is to examine the effect of replacing 
the Newtonian clock with a quantum clock $\{{\sigma}_t\}_{t\geq0}$, where the 
time that the quantum clock shows takes the form of a nonnegative increasing 
random process. 
We shall derive the dynamical equation satisfied by the state of 
a quantum system generated by a random unitary operator 
$\{{\hat U}({\sigma}_t)\}_{t\geq0}$, with a quantum clock $\{{\sigma}_t\}_{t\geq0}$ 
that ticks forward randomly in random amounts. 

The idea 
that the passage of time in quantum mechanics might arise from 
system-environment interactions has been envisaged by numerous authors 
\cite{PW,Lloyd,Marletto,Beige,Rost,Viotti}, but the literature typically stops short 
of explicitly constructing the ensemble dynamics of the quantum system in 
such models. In our model the exact nature of the dynamical equation depends 
on the statistical 
properties of the chosen clock. We can think of the clock process as modelling 
by proxy the effects of system's random interactions with its environment. Thus, 
in effect, we construct what might be called a `reduced form' model that 
encapsulates the effects of such interactions, hence enabling the theoretician to bypass, or at least temporarily put off, the task of modelling these interaction directly. Indeed, within our scheme one might classify the possible environmental effects via the specific choice of subordinator used as a basis for the model. There is an element of bootstrapping involved in 
such a construction since the Newtonian time still appears, but only as the 
ensemble average of the clock time registered by the members of a large 
ensemble of independent copies of the system. The ensemble is modelled by 
introducing a probability space  $\{\Omega, \mathcal F, \mathbb P\}$ on which 
both the state of the system and the quantum clock are random processes. 
The outcomes of chance in the sample space 
$\Omega$ correspond both to the individual elements 
of the ensemble and also to the specific trajectories of the random time process 
applicable to those elements. We require that for all $t\geq 0$ 
\begin{eqnarray}
{\mathbb E}[{\sigma}_t]=t \, , 
\label{expectation} 
\end{eqnarray} 
where ${\mathbb E}$ denotes the ensemble average, i.e. the Lebesgue integral 
over $\Omega$ taken with respect to the probability measure ${\mathbb P}$. 
This implies, in particular,  that $\sigma_0=0$ almost surely on $\Omega$, where 
time zero is interpreted as the moment the clock becomes operational, 
at which the state of the system faithfully reflects its initial 
preparation. 
Our requirement \eqref{expectation} then fixes the interpretation of the parameter 
$t$ as the ensemble average of the clock time shown for that value of the 
parameter.  We regard the specification of the probability space as implicit in the 
structure of the model, as usual in accordance with the standard theory of random 
processes. The relevant background in stochastic analysis can be found 
at  \cite{Sato, Applebaum}.  Then we can interpret the clock as a counting device 
that is synchronized with the frequency and magnitude of the system's interaction 
with its environment. In a generic environment we might expect the system to 
interact with environmental particles frequently by small magnitudes, but 
occasionally an environmental particle may have a greater impact. Then the 
counting `instrument' $\{{\sigma}_t\}_{t\geq0}$ will increase by very small amounts rather 
frequently, but will also increase by larger amounts occasionally. The point is that in 
reality the system registers the existence of an environment via the detection 
of particles, but such detection events are inevitably random both in 
Newtonian timing and in magnitude. 
It is also not unreasonable to suppose that the interactions of the system with 
its environment over disjoint intervals of the associated Newtonian time are 
independent. If these interactions are such that  our condition \eqref{expectation} 
holds, then we can interpret the Newtonian time $t$ as an emergent variable 
from an underlying quantum clock model. 

With these ideas in place, we show that the leading contribution to the dynamics 
of  the density matrix of the system is the von Neumann equation, regardless of 
the details of the statistical properties of the clock. However, the randomness of 
$\{{\sigma}_t\}_{t\geq0}$ will generate a modification of the von Neumann equation, the 
particulars of which depend on the statistics of the clock. Here we formulate the 
dynamical law of a quantum system subordinated by quantum clock constrained 
only by the requirements that we have indicated so far, subject to reasonable 
independence and homogeneity relations. 
Then we look in detail at the case for 
which the tick sizes follow a gamma distribution. This example fully illustrates the 
above-indicated feature of having a large number of very small ticks and a small 
number of much larger ticks, coordinated in such a way that condition 
\eqref{expectation} holds. 

The gamma model contains a parameter $\kappa$ that determines the balance of 
the rates at which jumps occur for jumps of different sizes.  The Newtonian time is 
recovered by taking the limit as $\kappa$ gets large. If the time shown by a 
quantum clock is viewed as a small perturbation away from the Newtonian time, 
then we expect $\kappa$ to be large, and hence we can expand the dynamics as 
a series in $\kappa^{-1}$. In this case the dynamical equation to leading order is 
given by an equation of the Lindblad type, where the Lindblad operator is the 
Hamiltonian. The model exhibits a generic structure for higher-order corrections 
that generalizes  both the von Neumann equation and the Lindblad term. 

The notion of a quantum clock introduced here does not directly concern quantum 
devices that can be used to estimate Newtonian time, nor the accuracy limits of 
such devices, for which there is an extended literature 
\cite{BH0,Oppenheim,Renner,Huber,Mitchison}. Rather, we focus on (a) the 
question on what is `time' in quantum mechanics, which we regard as a weighted 
count of the system-environment interaction, and (b) how Newtonian time can 
arise from this counting device.  Although our model is based on an underlying 
unitary evolution, the dynamics of the physical state is not time-reversal symmetric 
under Newtonian time. In our picture 
the fundamental breakdown in the unitarity of the evolution of the density matrix 
arises when we average over the statistics of the random clock.

A natural question to ask concerning any model for a quantum clock is whether it 
is consistent with the observed accuracy of atomic clocks. We find that the known 
accuracy of atomic clocks gives rise to a lower bound on the parameter $\kappa$. 
Taking the case of a ${^{133} \rm Cs^+}$ ion as an example, we show that 
$\kappa>10^{19}$ Hz is sufficient to ensure the model remains consistent with 
empirical observation. This bound on $\kappa$ can be made tighter by several 
orders of magnitude with an atomic clock of higher precision; however, it remains 
some twenty orders of magnitude smaller than Planck's frequency of $10^{43}$ 
Hz. Hence the Newtonian time can be replaced by a quantum clock for which both 
the tick sizes and their frequencies of occurrence range far away from the Planck 
scale. We conclude by arguing that the observation of small errors in atomic clocks 
would indicate that our approach might be a viable alternative to standard quantum 
theory equipped with the usual background Newtonian time. 

We begin by fixing a filtered probability space that supports a nondecreasing clock 
process $\{{\sigma}_t\}_{t\geq0}$, with the physical dimensions of time. As 
usual in the theory of jump processes we require that $\{{\sigma}_t\}_{t\geq0}$ 
should be right continuous for $t\geq 0$ with left limits on $t>0$ (the c\`adl\`ag 
property). Then for each $t>0$ we set  ${\sigma}_{t^-} = \lim_{s\uparrow t} 
{\sigma}_{s}$ and we define $\Delta {\sigma}_t = {\sigma}_{t} - {\sigma}_{t^-} $. 
It follows for all $t \geq 0$ that 
\begin{eqnarray}
{\sigma}_t = \sum_{0 \leq s \leq t} \Delta {\sigma}_s 
\end{eqnarray} 
and  we can hence interpret $\Delta {\sigma}_t$ as the size of the tick of the 
clock at the ambient Newtonian time $t$. The quantum system is modelled by an ensemble 
of states, the dynamics of which we represent by a random state process 
$\{\hat{\rho}_{t}\}_{t\geq 0}$. Thus, each element $\omega \in \Omega$ 
represents an element of the ensemble. Since $\Omega$ is in general 
uncountable, we are dealing with an abstraction here that is a little more 
general than the usual intuitive notion of an ensemble. We write ${\hat \mu}_t 
= {\mathbb E}[ \hat{\rho}_{t}]$ for the ensemble average, which we take to 
be the physical density matrix at $t$. 

We shall make three modelling assumptions. First, that $\{{\sigma}_t\}_{t\geq0}$ 
is a subordinator, that is, a nondecreasing L\'evy process.  The intuition behind 
this is that the environmental influences on the state over separate passages of 
Newtonian time are independent, and that the nature of these influences is 
unchanging over time. To this we add the Newtonian meantime expectation 
constraint $\eqref{expectation}$. 

Our second assumption is that the random state of the system can be modelled 
as a random unitary transformation of the initial state by setting 
\begin{eqnarray} 
{\hat{\rho}}_t = \re^{-{\rm i}{\hat H} {\sigma}_t} \,
\hat{\rho}_{0}\, \re^{{\rm i}{\hat H} {\sigma}_t},
\label{dynamics} 
\end{eqnarray}  
where $\hat H$ denotes the Hamiltonian. Here, for convenience we have set 
$\hbar = 1$. 
Our final assumption is that the clock process $\{{\sigma}_t\}_{t\geq0}$ and the 
initial state $\hat{\rho}_{0}$ are independent. 
The physical significance of this assumption is as follows. The initial state 
$\hat{\rho}_{0}$ can be taken to be pure or mixed and in either case can be taken to be deterministic or random.   The 
assumption that $\hat{\rho}_{0}$ and $\{{\sigma}_t\}_{t\geq0}$ are independent 
amounts to the requirement that the initial state is independent of the noisy 
environment into which it will be immersed. 

By a L\'evy process on $\{\Omega, \mathcal F, \mathbb P\}$ we mean a 
real-valued process $\{X_t\}_{t\geq0}$ that has stationary, independent 
increments such that 
\begin{eqnarray} 
\lim_{s\to t} \mathbb P[ \,| X_t - X_s | > \epsilon\,] = 0
\end{eqnarray} 
for every $t \geq 0$ and $\epsilon > 0$, with the property that there exists an 
$\Omega_0 \in \mathcal F$ with  $\mathbb P[\Omega_0] = 1$ such that for every 
$\omega \in \Omega_0$ it holds that $X_t(\omega)$ is c\`adl\`ag. We say that a 
L\'evy process is  increasing if the set of $\omega \in \Omega$ such that 
$X_t(\omega)$ is nondecreasing as a function of $t$ has probability one.  

An increasing L\'evy process is called a subordinator. In what follows we make use of the 
L\'evy-Khintchine formula, which gives an expression for the Laplace transform 
of such a process. Specfically, if $\{X_t\}_{t \geq 0}$ is a subordinator,  there 
exist a constant $\beta \geq 0$ and a measure $\nu({\rm d}x)$ on 
${\mathds R}^+$ satisfying  
\begin{eqnarray} 
\int_{[0,\infty)}  (1 \wedge x)\, \nu({\rd}x) < \infty,
\label{Levy condition} 
\end{eqnarray} 
such that for any complex $w$ with ${\rm Re}[w]\leq 0$ it holds that
\begin{eqnarray} 
\mathbb E \left[ {\re}^{ w X_t } \right] =  {\re}^{  \Psi(w) t }, 
\label{Levy-Khintchine} 
\end{eqnarray} 
with a Laplace exponent that can be expressed as a Lebesgue integral of the form
\begin{eqnarray} 
 \Psi(w) = \beta w + \int_{[0,\infty)} ( {\re}^{ w x } - 1 )\, \nu({\rm d}x).
\label{Levy exponent} 
\end{eqnarray} 
The L\'evy measure $\nu({\rm d}x)$ has the interpretation that for $0<a\leq b$ the integral
\begin{eqnarray} 
\nu[a,b] = \int_{[a,b]} \nu({\rm d}x)
\end{eqnarray} 
gives the average rate at which jumps in the range $[a,b]$ occur. 

The 
processes with which we are concerned fall into two classes. Those for which 
$\lim_{a\to 0} \nu[a,b] = \infty$ for fixed $b$ are said to have infinite activity. 
Otherwise, they have finite activity. A finite activity subordinator takes the form of 
a compound Poisson process with a nonnegative drift. The gamma process, 
which we consider shortly, is an example of a process of infinite activity. The 
constant $\beta$ can be chosen to fix the drift of $\{X_t\}_{t \geq 0}$. In particular, 
it follows from \eqref{Levy-Khintchine} and \eqref{Levy exponent}, along with a 
calculation, that \eqref{expectation} holds if and only if 
\begin{eqnarray} 
\beta  = 1 - \int_{[0,\infty)}  x \, \nu({\rd}x).
\label{drift condition} 
\end{eqnarray} 
Then in our model we use the notation $\{{\sigma}_t\}_{t\geq0}$ to denote a 
subordinator that satisfies the Newtonian meantime expectation constraint 
\eqref{expectation}. 

With these definitions in place we are in a position to work out the dynamics of 
the physical state of a system for which the associated ensemble takes  the form 
\eqref{dynamics}. Taking the matrix elements of such a state with respect to an 
energy basis, we obtain 
\begin{eqnarray} 
\langle E_j |{\hat{\rho}}_t| E_k \rangle =
\langle E_j |\hat{\rho}_{0}|E_k \rangle  \, \re^{{-\rm i}({E_j - E_k}) {\sigma}_t}.
\label{matrix_elements}
\end{eqnarray} 
For simplicity, we show the case of a nondegenerate 
Hamiltonian with a discrete spectrum. 
Then if we set 
\begin{eqnarray} 
\mu_{jk}(t) = \mathbb E\left [ \langle E_j |{\hat{\rho}}_t| E_k \rangle \right] 
\end{eqnarray} 
for the matrix elements of the density matrix in the energy basis, it follows from 
\eqref{Levy-Khintchine}, \eqref{Levy exponent}, and \eqref{matrix_elements}, 
when $\beta$ is given by \eqref{drift condition}, that
\begin{eqnarray} 
\mu_{jk}(t) = \mu_{jk}(0) \exp \left[ t \Big(\! - {\rm i}  (E_j - E_k) 
+  \int_{[0,\infty)} \left( {\re}^{ - {\rm i} (E_j - E_k) x  } + {\rm i}  
(E_j - E_k)x - 1 \right) \nu({\rm d}x)  \Big) \right].\,\,
\end{eqnarray} 
For the dynamics of the matrix elements it follows that
\begin{eqnarray}
\label{matrix_dynamics}
\frac{\rd{\mu}_{jk}(t)}{\rd t} &=&  \Big( \!- {\rm i}  (E_j - E_k)  +  \int_{[0,\infty)} 
( {\rm e}^{ - {\rm i} (E_j - E_k) x  }   
+ {\rm i}  (E_j - E_k)x - 1 ) \, \nu({\rm d}x) \Big)  {\mu}_{jk}(t)\, .
\end{eqnarray} 
Reinstating the Hamiltonian operator we then obtain 
\begin{eqnarray}
\label{dynamical_equation}
\frac{\rd{\hat \mu}_t}{\rd t} &=&  
- {\rm i}  [\hat H, {\hat \mu}_t]   
+  \int_{[0,\infty)} \left( {\rm e}^{ - {\rm i} {\hat H} x} {\hat \mu}_t \,{\rm e}^{  {\rm i} 
{\hat H} x}  +{\rm i}  [\hat H, {\hat \mu}_t]x -  {\hat \mu}_t  \right)  \nu({\rm d}x) , 
\label{general dynamical equation} 
\end{eqnarray} 
and this is the general form for the dynamical equation of the density matrix 
in a quantum clock model. 

One observes, by expanding the integrand of the L\'evy integral  as a power
series in $x$, that the zeroth and first order terms in $x$ in the integrand cancel 
and that the second order term gives a Lindbladian with an overall constant 
coefficient proportional to the second moment of the L\'evy measure, providing 
that this moment is finite. The higher order terms can also be worked out in the 
form of a series expansion if the L\'evy measure admits exponential moments -- 
that is to say, if there exists a strictly positive number $a$ such that 
\begin{eqnarray} 
 \int_{[1,\infty)} {\rm e}^{ a x }  \, \nu({\rm d}x) < \infty.
\label{exponential moments condition 1} 
\end{eqnarray}  
A necessary and sufficient condition for \eqref{exponential moments condition 1} 
to hold is that 
\begin{eqnarray} 
{\mathbb E}\! \left[ {\rm e}^{ a \sigma_t } \right] < \infty 
\label{exponential moments condition 2} 
\end{eqnarray}  
for some $t>0$, or equivalently, for every $t>0$. See Sato \cite{Sato}, 
Theorem 25.17. Such conditions hold, for example, for both the gamma process 
and the inverse Gaussian process considered below. 
The first few terms in the dynamical equation of the physical density matrix in 
that case are given  by 
\begin{eqnarray} 
\frac{\rd{\hat\mu}_t}{\rd t} &=& -\ri [{\hat H},{\hat\rho}_t] + c_2 \left( 
{\hat H}{\hat\mu}_t{\hat H}-\half({\hat\mu}_t{\hat H}^2 + {\hat H}^2{\hat\mu}_t )
\right)   + c_3 \left( 
\frac{1}{6} \ri [{\hat H}^3,{\hat\mu}_t]  -\frac{1}{2} \ri [ {\hat H}, {\hat H}
{\hat\mu}_t{\hat H}] \right)  \nonumber \\ && + c_4 \left(  \frac{1}{4} 
{\hat H}^2{\hat\mu}_t{\hat H}^2 - \frac{1}{6}({\hat H}{\hat\mu}_t{\hat H}^3 + 
{\hat H}^3{\hat\mu}_t{\hat H}) 
+ \frac{1}{24} ({\hat\mu}_t{\hat H}^4+{\hat H}^4{\hat\mu}_t) \right) 
+ \cdots , 
\label{eq:first few terms} 
\end{eqnarray} 
where we set 
\begin{eqnarray}
c_n = \int_{[0,\infty)} x^n \, \nu({\rm d}x) 
\end{eqnarray} 
for $n\geq 2$. The moments of the L\'evy measure can then be expressed in terms 
of the cumulants of ${\sigma}_t$ under $\mathbb P$ by use of \eqref{Levy-Khintchine}. 

One interesting result that follows immediately from the dynamical equation 
\eqref{general dynamical equation}  is a general tendency towards decoherence of the physical density matrix in the energy basis. 
It is easy to check first that the diagonal components (in the energy basis) of the random state process
$\{{\hat \rho}_t\}_{t\geq 0}$ are constant under the random unitary dynamics of \eqref{dynamics}, 
which shows that there is no state reduction to an energy eigenstate under such dynamics. 

On the other 
hand, it follows by a calculation using \eqref{matrix_dynamics} that for any 
off-diagonal component of the physical density matrix elements we have 
\begin{eqnarray}
|\mu_{jk}(t)|
= |\mu_{jk}(0)| \, \re^{-G_{jk} t} \, , 
\end{eqnarray} 
where for the rate of decoherence we find that
\begin{eqnarray}
G_{jk} = \int_{[0,\infty)} [ 1 - \cos( (E_j - E_k)x) ] \, \nu({\rm d}x).
\label{decoherence_rate}
\end{eqnarray} 
Then we observe that $G_{jk}$ is {\it strictly positive} for all $j\neq k$ providing that 
the spectrum of the L\'evy measure is not concentrated entirely on the 
characteristic time intervals associated with the spectrum of the Hamiltonian. 
Specifically, we require that the L\'evy measure is neither `tuned' to take the 
form 
\begin{eqnarray}
\nu({\rm d}x) = \frac{1}{\tau}\, \delta_{\tau}({\rm d}x) \quad {\rm with} \quad  \tau = \frac{2\pi n} {E_j - E_k},
\end{eqnarray}
for some $j\neq k$ and some integer $n \geq 0$, 
nor given by a sum of such expressions. Here $\delta_{x_0}({\rm d}x)$ denotes 
the Dirac measure concentrated at $x_0$. Then the physical density matrix 
diagonalizes in the energy basis at an exponential rate. 

To illustrate properties of the $\{c_n\}$ we consider a specific example 
of a quantum clock, which one might call a `gamma clock'. In this case 
$\{{\sigma}_t\}_{t\geq0}$ follows a gamma process, for which the increments 
are gamma distributed and the probability density is given by
\begin{eqnarray}
\mathbb P\left[{\sigma}_t \in {\rm d} x \right] =\frac{1} {{\Gamma}(\kappa t)}
\kappa^{\kappa t}\,x^{\kappa t-1}\,\re^{-\kappa x} \,  {\rm d} x,  
\label{gamma density} 
\end{eqnarray} 
where ${\Gamma(z)}$, $z>0$, denotes the gamma function.
It follows that ${\mathbb E}[{\sigma}_t]=t$, as required, and that 
${\mathbb E}[\left({\sigma}_t - t \right)^2]=\kappa^{-1}t$. 
Such a process is known to model the water level in a dam 
and also has numerous applications in risk theory, finance, and insurance 
\cite{Kendall, Dufresne1991,Yor2007,BHM}. The associated L\'evy measure 
vanishes for $x\leq 0$ and is given by
\begin{eqnarray}
\nu({\rm d}x) = \kappa \,  
x^{-1} \,\re^{-\kappa x} \, {\rm d} x 
\label{gamma Levy measure} 
\end{eqnarray} 
for $x>0$. Then we have
\begin{eqnarray}
\nu[a,b] =  \kappa \left( {\rm E}_1( \kappa a) -  {\rm E}_1( \kappa b)\right).
\label{gamma jump rate} 
\end{eqnarray} 
Here, for the exponential integral function (incomplete 
gamma function) we have written 
\begin{eqnarray}
{\rm E}_1( z) = \int_z^{\infty} x^{-1} \re^{- x} \,{ \rm d} x.
\end{eqnarray} 
The gamma clock thus has the property 
that over any finite interval $\tau$ of Newtonian time the average number of 
jumps is $\nu[a,b] \,\tau$ for jump size in the range $[a,b]$ and hence is 
unbounded as $a$ approaches zero. 

\begin{figure}[t]
\centering
{\includegraphics[width=0.90\textwidth]{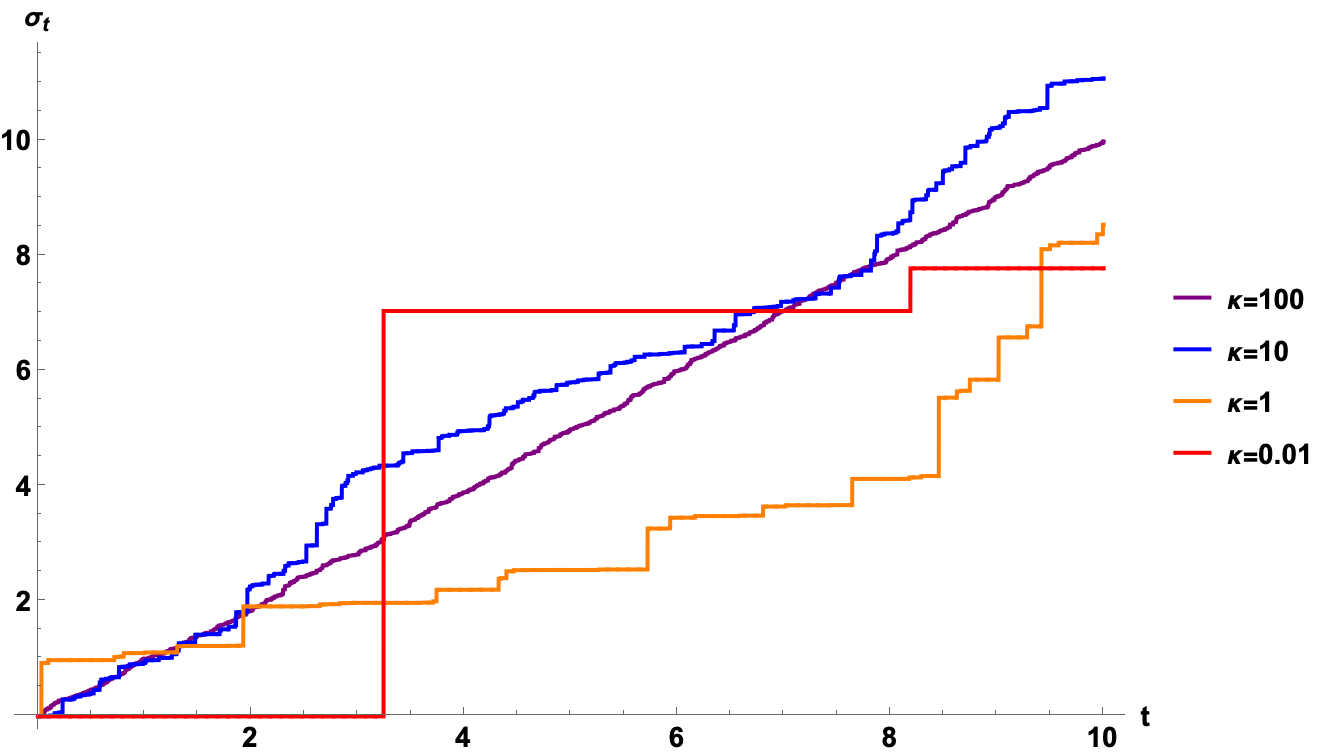}}\hfill
\caption{\textit{Gamma clock}. 
The behaviour of a gamma clock for different values of the rate parameter 
$\kappa$ is simulated 
 for $\kappa=100$, $\kappa=10$, $\kappa=1$ and $\kappa=0.01$. 
For larger $\kappa$ the 
clock time approaches the Newtonian time.  For small $\kappa$, visible ticks 
become rare, but the jump size can be significant. 
}
\label{Fig:0}
\end{figure}  

The parameter $\kappa > 0$ 
determines the rates at which ticks of various magnitudes occur \cite{HSB2020} 
and we recover the Newtonian 
time ${\sigma}_t\to t$ in the limit $\kappa\to \infty$. 
The moments of the gamma clock can be worked out 
by use of the moment generating  function
\begin{eqnarray}
{\mathbb E}\left[\re^{\alpha {\sigma}_t}\right] = 
\left( 1- \frac{\alpha}{\kappa}\right)^{-\kappa t},
\label{eq:MGF}
\end{eqnarray} 
which is defined for $\alpha < \kappa$. Examples of the behaviour of a gamma 
clock, as a function of Newtonian time, are 
shown in Figure~\ref{Fig:0} for various values of $\kappa$. 
From (\ref{eq:MGF}) we deduce that 
\begin{eqnarray}
{\mathbb E}[( {\sigma}_t)^n] = t  \left(\! t+\frac{1}{\kappa}\!\right)\!\! 
\left(\!  t+\frac{2}{\kappa}\!\right)\!\! \, \cdot \, \cdot \, \cdot \,
\left(\! t+\frac{n-1}{\kappa}\!\right),
\label{eq:11} 
\end{eqnarray}
and if we write $\lambda=\kappa^{-1}$ then for $n\geq2$ we find that 
$c_n = (n-1)!\, \lambda^{n-1}$.
Thus in the Newtonian  limit $\lambda\to0$ we recover 
the von Neumann equation of standard quantum mechanics, but if the quantum 
clock has a small deviation from the Newtonian time so that $\lambda\ll1$ 
but $\lambda>0$, then we get a small energy-driven decoherence effect. The 
largest correction to the von Neumann term is given in that case by 
the Lindblad term. Then there are further higher-order corrections that 
generalize the von Neumann equation for even powers in $\lambda$ and 
the energy-driven Lindblad equation for odd powers in $\lambda$. 

The properties of a gamma clock are typical among 
quantum clock models. Suppose, alternatively, that 
${\sigma}_t$ has an inverse-Gaussian distribution, given by the density
\begin{eqnarray}
\mathbb P\left[{\sigma}_t \in {\rm d} x \right]  = \sqrt{\frac{\kappa t^2}{2\pi x^3}} \, \exp\left[{-\frac{\kappa}{2x}
(x-t)^2} \right] .
\label{eq:6} 
\end{eqnarray}  
Such a distribution models the first passage time of a Brownian motion 
\cite{Wasan}. In this case, the associated moment generating function is  
\begin{eqnarray}
{\mathbb E}\left[\re^{ \alpha {\sigma}_t}\right] = 
\exp\left[ \frac{\kappa}{2}\left( 1-\sqrt{ 1-\frac{4
\alpha}{\kappa}}\right) t \right] , 
\label{eq:7} 
\end{eqnarray} 
and we deduce that 
\begin{eqnarray}
c_n = 2 \frac{(2n-3)!} {(n-2)!}\, \lambda^{n-1}
\end{eqnarray} 
for $n\geq2$. Though the $\{c_n\}$ differ, the qualitative features of the dynamics 
of the inverse-Gaussian clock are similar in many respects to those of a gamma 
clock. 

\begin{figure}[t]
\centering
{\includegraphics[width=0.90\textwidth]{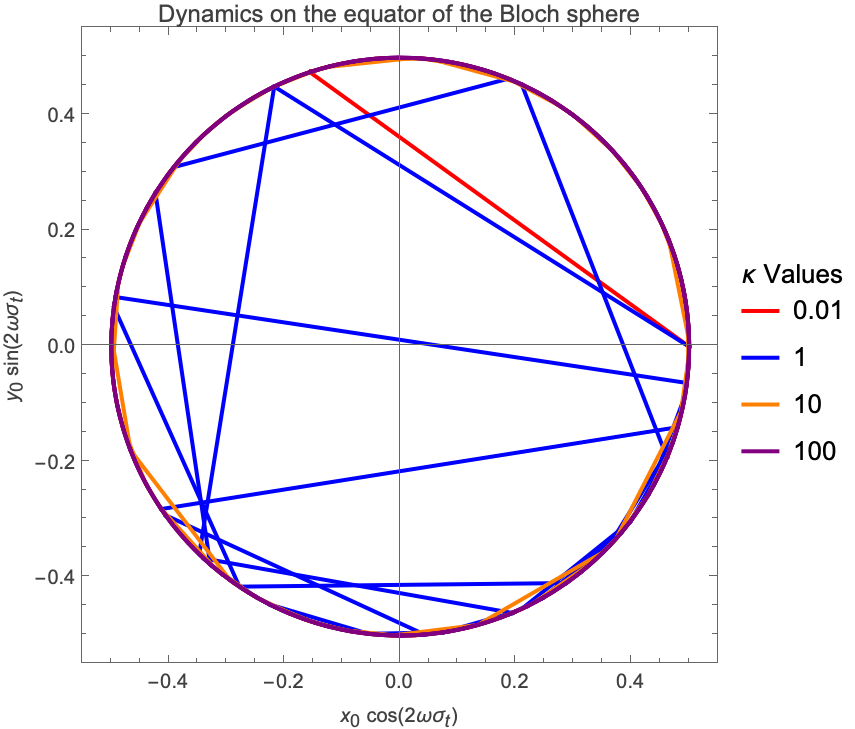}}\hfill
\caption{\textit{Quantum dynamics under a quantum clock}. 
Dynamical trajectories  associated with different 
jump rates are simulated here, when the initial state lies on the equator of 
the Bloch sphere. Since the state jumps from one point of the 
equator to another, to aid visualization these points are joined by straight lines, although lines are not part of the dynamics. 
}
\label{Fig:X}
\end{figure}  

As an example of a system with a quantum clock, consider a spin one-half particle 
immersed in a magnetic field in the $z$ orientation, with 
${\hat H}=\omega {\hat\sigma}_z$, under a gamma clock. The 
evolution of 
the quantum state is simulated in Figure~\ref{Fig:X} 
for an initial state that is an eigenstate of ${\hat\sigma}_x$. Since the 
dynamics for the random state $\{\hat{\rho}_t\}$ are unitary, an initial pure state 
on the surface of the Bloch sphere will remain on the latitudinal circle containing the 
initial state, which is the equator here. However, as the quantum 
clock jumps, the state also jumps along the circle. 

\begin{figure}[t]
\centering
\subfloat[Dynamical trajectories of ${\hat\mu}(t)$ for a small $\lambda$.]{\includegraphics[width=0.46\textwidth]{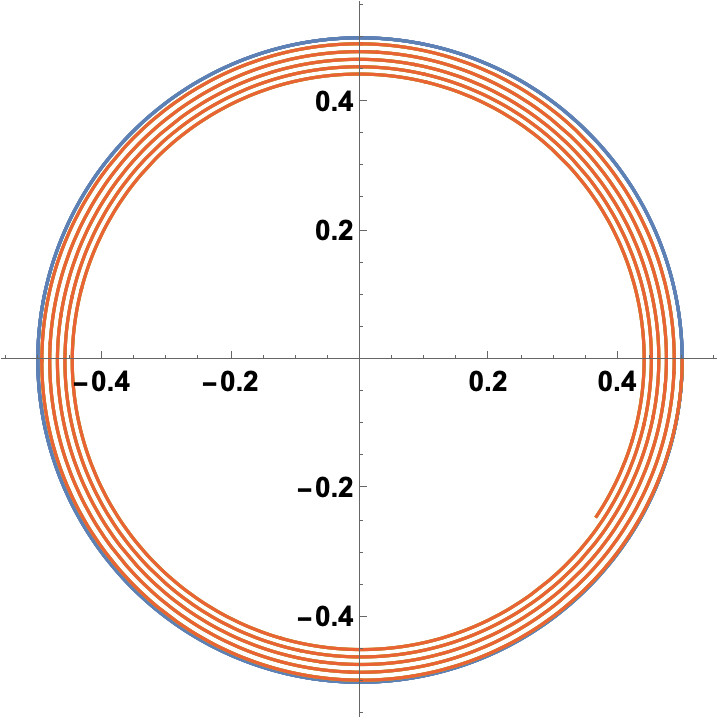}}\hfill
\subfloat[Dynamical trajectories of ${\hat\mu}(t)$ for a large $\lambda$.]{\includegraphics[width=0.46\textwidth]{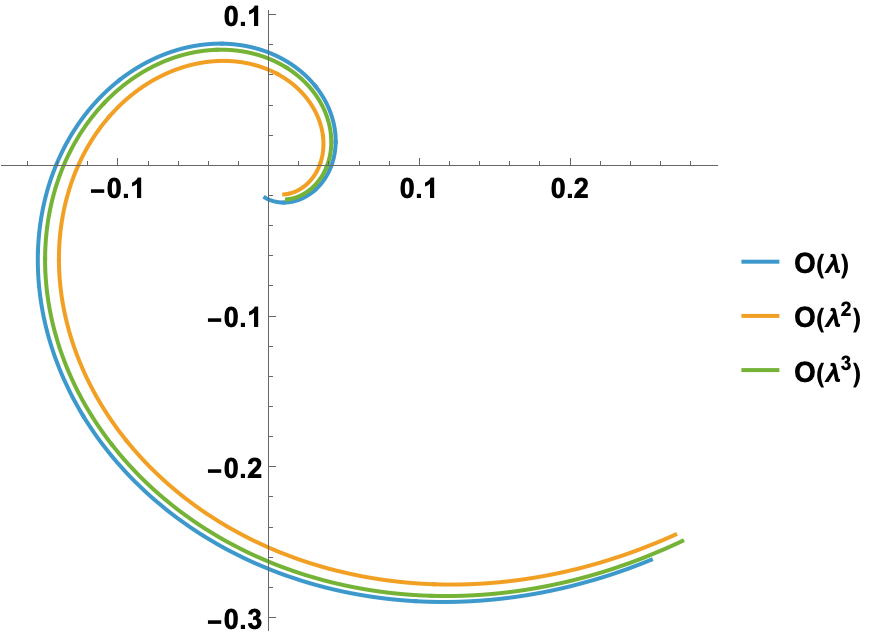}}\hfill
\caption{
\textit{Dynamics of the density matrix}. 
In the case of a spin-$\frac{1}{2}$ system with the Hamiltonian ${\hat H}=
\omega{\hat\sigma}_z$, the dynamical trajectory 
${\hat\mu}(t)$ for the density matrix is confined to a latitudinal disk 
inside the Bloch sphere. The resulting trajectories of order up to $\lambda^0$, 
$\lambda^1$, $\lambda^2$, and $\lambda^3$ are plotted here with the 
initial state given by the eigenstate of ${\hat\sigma}_x$ with the eigenvalue 
$+1$ so that the orbits lie on the equatorial disk. Here we set 
$\omega=0.8$ for the angular frequency. 
(a) For a small $\lambda$, set at $\lambda=0.005$ here, the 
deviation away from the leading Lindblad (of order $\lambda$) term is 
negligible. 
(b) For a larger $\lambda$, which is set at $\lambda=0.5$ here, the 
trajectories become more distinguishable. The overall 
exponential damping is slightly suppressed by including the $\lambda^3$ 
term, as compared to the trajectory truncated at $\lambda^2$. Here, the 
unitary ($\lambda=0$) orbit is omitted. 
}
\label{Fig:1}
\end{figure}

The corresponding dynamics for the physical density matrix
${\hat\mu}_t$ in (\ref{eq:first few terms}) can
be solved exactly to any order in $\lambda$ in this example.
Unlike ${\hat\rho}_t$, which evolves unitarily, so the purity of the state
is conserved, the evolution of the ${\hat\mu}_t$ takes place
inside the Bloch sphere without the preservation of purity. 
The resulting orbits of order up to $\lambda^0$ (the unitary case),
up to $\lambda^1$ (unitary dynamics plus Lindblad term), up to $\lambda^2$
(including the modified unitary term), and up to $\lambda^3$ (also including
the generalized
Lindblad term) are plotted in Figure~\ref{Fig:1}(a). For
this choice of Hamiltonian, the orbits lie on a latitudinal disc inside the Bloch
sphere containing the initial state. 
Based on the parameter values chosen, all orbits essentially coincide,
with the exception of the unitary orbit corresponding to the $\lambda^0$ term.
When $\lambda$ is increased, the orbits corresponding to the
truncation of the dynamical equation with different powers in $\lambda$ become
visibly distinguishable, as shown in Figure~\ref{Fig:1}(b).

We turn to explore whether a potential detectable signal of a quantum clock 
model might be identified. To this end we consider {\em the spread of 
the wave packet} for a free particle evolving with a quantum clock. In fact, such a system can be given a more or less exact treatment for any choice of the L\'evy subordinator. 

For simplicity we look at the one-dimensional situation, letting $\hat p$ and $\hat q$ denote the momentum and position operators, respectively, and for the Hamiltonian we write
\begin{eqnarray} 
{\hat H}= \frac{1}{2m} {\hat p}^2
\end{eqnarray} 
where $m$ denotes the mass. We are interested in the behaviour of the 
spread, that is, the unconditional variance (or squared `uncertainty') of the position of the particle, which is given by the following expression: 
\begin{eqnarray} 
{\rm Var}_t (\hat q) ={\mathbb E}\left[ {\rm tr}({\hat q}^2{\hat\rho}_t) \right] - 
\left( {\mathbb E}[{\rm tr}({\hat q}{\hat\rho}_t)]\right)^2 .
\label{eq:totalvariance} 
\end{eqnarray} 

We assume that the ${\hat\rho}_t$ evolves unitarily under a 
quantum clock in accordance with (\ref{dynamics}). To work out ${\rm Var}_t (\hat q)$ we 
make use of the relation 
\begin{eqnarray}
\re^{{\rm i}\sigma_t{\hat p}^2/2\hbar m}\,{\hat q}\,\re^{-{\rm i}\sigma_t{\hat p}^2/2\hbar m} = 
  {\hat q} + \frac{\sigma_t}{m}\,{\hat p}  \,,  
\end{eqnarray} 
which follows from the commutation relations for $\hat p$ and $\hat q$. A calculation then shows that 
\begin{eqnarray}
{\rm tr}({\hat q}{\hat\rho}_t) = {\rm tr}({\hat q}{\hat\rho}_0) + 
\frac{\sigma_t}{m} \,  {\rm tr}({\hat p}{\hat\rho}_0), 
\end{eqnarray} 
where ${\hat\rho}_0$ is the initial state, and that
\begin{eqnarray}
{\rm tr}({\hat q}^2{\hat\rho}_t) = {\rm tr}({\hat q}^2{\hat\rho}_0) + 
\frac{2\sigma_t}{m}\, {\rm tr}\left( \frac{{\hat p}{\hat q}+{\hat q}{\hat p}}{2} \, 
{\hat\rho}_0\right) + \frac{\sigma_t^2}{m^2} \,{\rm tr}({\hat p}^2{\hat\rho}_0) \, .
\end{eqnarray} 
To take expectations of these expressions over the clock statistics 
we remark that 
\begin{eqnarray}
{\mathbb E}[\sigma_t] = \Psi'(0) \, t \quad {\rm and} \quad  
{\mathbb E}[\sigma_t^2] = \Psi''(0) \, t + (\Psi'(0))^2 t^2 \, ,
\end{eqnarray} 
where $\Psi(w)$ is the L\'evy exponent defined in (\ref{Levy-Khintchine}).   
It then follows from (\ref{eq:totalvariance}) that 
\begin{eqnarray}
{\rm Var}_t (\hat q) = {\rm Var}_0 (\hat q) + 2 \left( 
\frac{\Psi'(0)}{m}\, {\rm Cov}_0[{\hat q},{\hat p}] + \frac{\Psi''(0)}{m} 
\, {\langle \hat H  \rangle}_0 \right) t +  \frac{(\Psi'(0))^2}{m^2}\, {\rm Var}_0 (\hat p) \, t^2 \, . 
\label{eq:Ehrenfest2} 
\end{eqnarray} 
Here ${\rm Cov}_0[{\hat q},{\hat p}]$ is the covariance of ${\hat q}$ and 
${\hat p}$ under ${\hat\mu}_0$, ${\rm Var}_0 (\hat p)$ is the variance of 
${\hat p}$ under ${\hat\mu}_0$, and ${\langle \hat H  \rangle}_0$ is the 
expectation value of the energy under ${\hat\mu}_0$. Note 
that our postulate for $\{\sigma_t\}$ implies that $\Psi'(0)=1$. 

If, in addition, 
$\Psi''(0)=0$, which holds if and only if $\{\sigma_t\}$ is Newtonian time 
up to an additive constant, 
then (\ref{eq:Ehrenfest2}) reduces to a well-known result that follows 
from the Ehrenfest theorem for a free-particle system. However, for a 
generic quantum clock we have $\Psi''(0)\neq0$. Therefore, the term 
involving the energy in \eqref{eq:Ehrenfest2} predicts a distinctive 
deviation from standard quantum mechanics in the case of a quantum clock.

Next, we propose to determine conditions under which the decoherence 
effects induced by a quantum clock model are consistent 
with the precision limits imposed by atomic clocks. To this end, we look again at
the gamma clock, for which the elements of the density matrix 
in the energy basis are given by  
\begin{eqnarray}
\mu_{jk}(t) = \mu_{jk}(0) \left( 1-\ri \, \frac {\nu_{jk} } {\kappa}  \right)^{-\kappa t} \, ,
\end{eqnarray} 
where $\nu_{jk}=E_j-E_k$. The decoherence 
rate $G_{jk}$ for $\mu_{jk}(t)$ for large $\kappa$ is therefore 
\begin{eqnarray}
G_{jk} =  \frac{\kappa}{2} \log\left( 1+
\frac{\nu_{jk}^2}{\kappa^2}\right) \approx  \frac{\nu_{jk}^2}{2\kappa} \, . 
\label{eq:16} 
\end{eqnarray} 
Following the analysis of Weinberg \cite{weinberg2016} taken together with that of  \cite{BH}, 
we can place a lower bound on $\kappa$ by consideration of the accuracy of an atomic clock. Weinberg argues, 
in particular, that the decoherence rate must satisfy a bound of the form 
$G_{jk}\,T_R < 1$, where $T_R$ is the Ramsey time associated with the clock 
\cite{weinberg2016}. $T_R$ varies with the type of clock, 
but is typically of the order of a  few seconds. For the energy gap $\nu_{jk}$ 
in (\ref{eq:16}), we consider the ${^{133} \rm Cs^+}$ ion, for which the 
hyperfine transition frequency is known to great accuracy, given by 
$9.192\,631\,77$ GHz. The corresponding energy 
difference is. 
\begin{eqnarray}
\Delta \nu_{\rm Cs} \approx 3.801 \times 10^{-5} \,{\rm eV}.
\end{eqnarray} 
To put a bound on $\kappa$ we reinstate Planck's 
constant, so 
\begin{eqnarray}
G_{jk}=\frac{1}{2\hbar^2 \kappa} ({\Delta \nu_{\rm Cs}})^2. 
\end{eqnarray} 
Then $G_{jk}<1\,{\rm s}^{-1}$ implies
$\kappa>10^{19}\,{\rm s}^{-1}$. 
That is to say, provided the jump rate parameter is such that 
$\kappa>10^{19}\,{\rm s}^{-1}$, the gamma clock will respect the 
precision bound imposed by the atomic clock. But how do the tick sizes of 
such a clock compare to the timescale set by Planck's time? 

For any $\delta>0$ the probability of the 
clock ticking exactly $n$ times by at least the amount  $\delta$ over a 
Newtonian time interval of length $\tau$ is given by 
\begin{eqnarray}
{\mathbb P}(n)=\frac{1}{n!} \,(r_\delta \tau)^n \, \re^{-r_\delta \tau},
\end{eqnarray} 
where $r_\delta$ is the rate at which jumps of size $\delta$ or more occur.  
However, by \eqref{gamma jump rate} we have 
\begin{eqnarray}
r_\delta =  \kappa \,\rm E_1( \kappa \delta).
\label{r}
\end{eqnarray} 
Thus, from the Poisson law we see that the probability of the clock ticking at 
least once by an amount greater than or equal to $\delta$ in the Newtonian 
interval $\tau$ is $1-{\mathbb P}(0)=1-\re^{-r_\delta \tau}$, where $r_\delta$ 
is given by \eqref{r}. If $\delta$ is taken to be the Planck time 
$5.39\times 10^{-44}\,{\rm s}$ and we set $\kappa=10^{19}\,{\rm s}^{-1}$, then 
one finds that $r_\delta=5.53\times10^{20}\,{\rm s}^{-1}$. Therefore, the 
probability of the quantum clock ticking by an amount no less than the Planck 
time over, say, 247 zeptoseconds, $\tau=2.47\times 10^{-19}{\rm s}$, the 
smallest measurable time interval, the time it takes for light to travel across a 
molecule of hydrogen \cite{GRUNDMANN}, is for all practical purposes
$100\%$. That is, over the course of the shortest measurable time 
interval, a quantum clock will have jumped at least once by an amount greater 
than the Planck time with near certainty.  

In a quantum clock model, the estimation error for the Newtonian time $t$ is 
bounded below by the inverse Fisher information \cite{BH0}. If the tick size of 
the clock is distributed according to (\ref{gamma density}), then a calculation 
shows that the Fisher information for the parameter $t$ is given by 
$\kappa^2 \psi'(\kappa t)$, where $\psi(u)=\rd \ln\Gamma(u)/\rd u$ is the 
digamma function and $\psi'(u)$ its derivative. It follows that for 
$\kappa=10^{19}\,{\rm s}^{-1}$ the estimation error bound for $t$ 
given by the inverse Fisher information 
is a tenth of a second even 30 billion years after the initial preparation of the 
system. This error lower bound is 
consistent with the accuracy limit of an optical atomic clock 
with an error of about a second over 30 billion years \cite{Aeppli}. 

Let us comment on the implications of the estimate 
$\kappa=10^{19}\,{\rm s}^{-1}$ for the consequences of the extended 
Ehrenfest theorem 
(\ref{eq:Ehrenfest2}). Specifically, for a gamma clock we have 
$\Psi''(0)=\kappa^{-1}$, 
so for a large $\kappa$ the correction term is suppressed. However, 
for a particle with a high average velocity -- for instance a high-energy 
electron beam inside a long vacuum accelerator for which the velocity 
is close to the speed of light -- the modified dispersion effect predicted by a 
quantum clock model is potentially testable. 

The idea that time may be discrete at or below the Planck time scale is often 
entertained in the literature. Our analysis shows that the discreteness of time 
can arise at a larger time scale without being inconsistent with empirical 
properties of atomic clocks. The novelty in our model is the notion that the both 
occurrences of the ticks and the magnitudes of the tick size of the clock are 
random. As a consequence, Newtonian time can be seen to emerge from a 
statistical averaging over an ensemble of quantum clocks. Although the 
dynamical evolution of the quantum state is unitary, the counting aspect of a 
quantum clock makes it time-irreversible from the perspective of Newtonian time. 
The idea of a time-irreversible random unitary motion has been considered by 
Peres \cite{Peres} and Adler \cite{Adler2000} in a different context;
but the Peres-Adler ansatz is based on a Gaussian white noise \cite{Hida} to 
model system-environment interaction strengths, and this does not give rise to 
a quantum clock model. 
If the clock process $\{\sigma_t\}_{t \geq0}$ is of the special form 
$\sigma_t = \gamma^{-1} N_t$ where $\{N_t\}_{t \geq0}$ is a standard 
Poisson process with rate $\gamma >0$, then for the L\'evy measure of the 
subordinator one has $\nu(\rd x)=\gamma\, \delta_{\gamma^{-1}}({\rm d}x)$, 
in which  case \eqref{expectation} is satisfied and \eqref{dynamical_equation} 
reduces to Milburn's equation (2.7) of reference \cite{Milburn}. 
Hence, as a byproduct of our investigation, we obtain an exact derivation of 
Milburn's model for `intrinsic decoherence', avoiding some of the heuristics 
of Milburn's original approach. 

The accuracy of atomic clocks, with an error of a second over billions of years, 
imposes a lower bound on the jump rate parameter. The fact that an atomic 
clock admits a small error suggests that the quantum clock models 
might provide a feasible alternative to standard quantum theory, which is 
directly constructed on the presumption of an 
ambient Newtonian time. Our approach also assumes the existence of an ambient Newtonian time, at least implicitly. In particular, we have relied, as  the basis for our examples, on the use of the theory of L\'evy subordinators.  As is always the case in the theory of stochastic processes and their applications, such an analysis incorporates the existence of a deterministic, strictly increasing ``ambient" time. This is modelled by the introduction of a so-called filtration on the probability space in the form of a nested family of sigma-algebras. 

This well-established, very useful idea dates back many decades to foundational work appearing in the landmark texts of Doob (1953) and Meyer (1966). Our definition of a quantum clock makes use of precisely this structure in our introduction of a pure-jump subordinator, where the value taken by the process as it jumps plays the role of the observable time that drives the dynamics of the quantum state. The so-called Newtonian time (which is what we call the ambient time) is not directly observable but can be inferred by an averaging procedure in the case of an ensemble of independent systems. In this sense, it is an ``emergent" concept, though as always in such considerations, the rabbit goes into the hat at the beginning of the discussion, in the very definition of what one means by a stochastic process. It should not be surprising then that the emergent Newtonian time agrees with the ambient time implicit in the definition of the underlying counting process. The key point is that this is the only way to define a counting process. 

Though in this paper we have looked at temporally homogeneous 
clocks, where the statistics of the ticks do not depend on 
the Newtonian time, one can envisage a model in which the rate and 
distribution of the jumps is time 
inhomogeneous -- for instance, a model in which the jump rate is initially low (so 
the clock can tick by large amounts over a short period of Newtonian time), but 
then gradually increases. 
Such a model may find applications in the quest for a 
quantum theory of cosmology.

\vspace{0.5cm} 
\noindent {\bf Acknowledgements} 
\vspace{0.5cm} 

\noindent 
The authors wish to thank Eva-Maria Graefe and Levent A.~Meng\"ut\"urk for helpful discussions. 
DCB acknowledges support from EPSRC grant EP/X019926/1.

\end{document}